# A Coarsening Approach to the Troll Aquifer Model

David Landa-Marbán[1*], Tor H. Sandve[1], Sarah E. Gasda[1,2]

[1] Division of Energy and Technology, NORCE Research AS, Bergen, Norway
[2] Department of Physics and Technology, University of Bergen, Bergen, Norway

* Corresponding author e-mail: dmar@norceresearch.no

## Abstract

Effective management of pressure communication and interference between concurrent $CO_2$ storage operations is essential for the development of gigaton-scale storage hubs. Coarse models in reservoir simulation offer a simplified representation of the subsurface to efficiently predict pressure distribution. By averaging properties over larger grid blocks, coarsened models significantly reduce computational demands, enabling faster and more manageable simulations. The approach focuses on preserving key physical properties, namely cell connectivity (transmissibilities) and storage capacity (pore volumes). The effectiveness of the coarsened model is demonstrated by applying the improved well location derived from the coarsened model to the full-resolution Troll aquifer model, resulting in improved pressure distribution. The coarsened model enabled the execution of approximately 100,000 simulations over five days using a local server with 144 CPUs, an effort that would have required around seven months using the original model. The coarsening methodology is implemented using *pycopm*, an open-source tool design to tailor geological models based on industry-standard grid formats, facilitating its application to other models and applications in reservoir engineering.

## 1. Introduction

Geological carbons sequestration (GCS) has matured into a viable climate mitigation technology, where the critical next frontier is transitioning from individual megaton-scale projects to gigaton-scale storage hubs [1]. Achieving this scale requires the utilization of large regional aquifers, where it becomes essential to evaluate the cumulative pressure impact of multiple operators injecting $CO_2$ across many distinct but hydraulically-connected sites. Analysis of this system with respect to commercial targets and geomechanical constraints requires pressure modelling of the entire regional aquifer, often within multi-objective optimization workflows that involve extensive scenario-based simulations [2]. However, the computational demands of such workflows make it impractical to employ detailed multiphysics models at the regional scale. Therefore, scalable modelling strategies are required to balance physical fidelity with computational efficiency, enabling robust decision-making for large-scale GCS deployment.

Model reduction aims to simplify complex geological models while retaining the essential dynamics, thereby enabling faster and more efficient simulations to support decision-making processes such as uncertainty quantification, history matching, and optimization. The selection of an appropriate model reduction technique is influenced by several critical factors, including the specific objectives of the study, the complexity of the geological system, the availability and quality of the data, and the computational resources at hand. Examples in GCS of model reduction are vertical equilibrium [3] and subgrid-scale modelling [4].

The most efficient and appropriate model reduction strategy will seek to preserve only those flow physics that are most important for the problem of interest. For regional-scale pressure management in GCS, key considerations include the connected pore volume and boundary conditions of the aquifer, as well as reservoir connectivity characterized by transmissibility. Conversely, a coarse-scale representation of $CO_2$ migration would require additional focus on pseudo-relative permeability and capillary pressure. Consequently, coarsening is an appropriate model reduction technique, as it involves upscaling reservoir properties over large grid blocks while preserving the dominant flow characteristics.

As [5-7] points out, the coarsening needs to be tailored to the problem. Successful coarsening balances accuracy and efficiency and considers both the physics and the usage of the model. User-friendly and flexible open-source packages enables researchers and engineers to tailor models to their needs for advanced applications, such as optimization, machine learning, and data assimilation. Incorporating open-source code into coarsened reservoir models also enhances the reproducibility of published simulation results and supports transparent, collaborative research. By leveraging these capabilities, coarsened models can be continuously improved and tailored to evolving project needs, ultimately contributing to more efficient and robust reservoir management strategies.

The literature includes several notable contributions involving open-source code for model reduction. The MRST project offers a wide range of flexible coarsening algorithms as well as tools for upscaling [7]. Recently, data-driven flow network models have been developed [8, 9]. In these models, the flow is modelled as one-dimensional tubes between wells. These models are highly efficient but depend on access to a large set of data and are not directly applicable to typical $CO_2$ storage applications.

The development of *pycopm* [10] was motivated by the need for user-friendly, open-source tools for model coarsening. This tool is designed to work with the industry-standard input format for reservoir simulation decks and produces coarsened decks in the same format, ensuring compatibility with existing workflows. This distinguishes it from the packages available in MRST. The main functionalities provided by *pycopm* include grid coarsening, grid refinement, submodel extraction (sector models), and affine transformations such as scaling, rotation, and translation.

This paper presents the application of the coarsening functionality in *pycopm* to a publicly available static model of the Troll aquifer, recently released by the Norwegian Offshore Directorate. The study begins with an overview of the theoretical framework underlying the adopted coarsening approach. Next, the main characteristics of the Troll aquifer model are described. The focus then shifts to the coarsened version of the Troll model, which is used in an intensive workflow aimed at improving well placement to achieve a more uniform reservoir pressure distribution. Finally, the results obtained from both the Troll aquifer model and coarsened version are discussed.

## 2. Methodology

The following subsections first describe the theoretical background of the coarsening approach, followed by an introduction to the Troll aquifer model.

Note that the developed coarsening methodology is openly available through an online repository at https://github.com/cssr-tools/pycopm, which includes extensive documentation and usage examples. This user-friendly tool supports the industry-standard input format for reservoir simulation decks and generates output files with grid and property data in the same format. In addition to coarsening, *pycopm* provides functionality for grid refinement, sector model extraction, and affine transformations.

This study also employs the OPM Flow open-source reservoir simulation software to demonstrate the coarsening methodology. OPM Flow also supports industry-standard input and output formats [11]. It includes capabilities for simulating GCS [12], and its implementation has been successfully benchmarked against other simulators [13].

### 2.1 Grid coarsening

Corner-point grids are widely used in subsurface simulations due to their flexibility in representing complex geological features, such as faults. These grids are defined by vertical pillars and horizontal lines connecting the pillars, forming cells that typically have six faces (hexahedrons). In addition, corner-point grids support the definition of connections between non-neighboring cells, known as non-neighboring connections (NNCs). A specific case of corner-point grids is the Cartesian regular grid, which features uniform cell geometry. These grids may also include inactive cells, which are defined as cells with zero pore volume and no connectivity to active cells. Figure 1 illustrates an example of a corner-point grid containing faults and inactive cells.

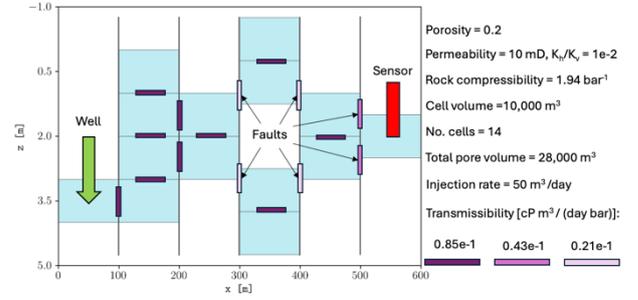

Figure 1: Example of a corner-point grid with non-neighboring connections, resulting in varying transmissibility values across cell faces.

We define the grid coarsening operator $X_{xyz}^{\mathcal{C}}$, which transforms an input grid $\Omega$ into a coarsened grid $\Omega^*$:

$$X_{xyz}^{\mathcal{C}} (\Omega) \rightarrow \Omega^*.$$

In this notation, the asterisk denotes coarsened quantities, while the superscript $\mathcal{C}$ in the operator highlights the specific approach used for cell clustering. The coarsening process involves the removal of selected vertical pillars and horizontal lines from the original grid.

To provide flexibility in the coarsening process, cell clustering can be specified using arrays in the $x$, $y$, and $z$ directions, which define the vertical pillars and horizontal lines to be removed. An important consideration in this process is how to treat inactive cells within a cluster. To address this, three options are implemented for determining the activity status of coarsened cells: min, max, and mode. The mode option is used by default, meaning that a coarsened cell is considered active if most cells in the cluster are active.

Naturally, the pore volume in a coarsened cell $\Phi_{i^*,j^*,k^*}^*$ ($i^*$, $j^*$, and $k^*$ referring to the cell coarsened indices in the $x$, $y$, and $z$ direction respectively) is equal to the sum of pore volume from the corresponding cells $\Phi_{i,j,k}$ in the original model, which are part of the cluster $\mathfrak{C}_{i^*,j^*,k^*}$:

$$\Phi_{i^*,j^*,k^*}^* = \sum_{(i,j,k) \in \mathfrak{C}_{i^*,j^*,k^*}} \Phi_{i,j,k}.$$

Based on this definition, the porosity in the coarsened model $\phi_{i^*,j^*,k^*}^*$ can be directly computed as:

$$\phi^*_{i^*,j^*,k^*} = \frac{\Phi^*_{i^*,j^*,k^*}}{\mathbb{V}^*_{i^*,j^*,k^*}}$$

where $\mathbb{V}^*_{i^*,j^*,k^*}$ is the geometric volume of the coarsened cell.

Several upscaling methods are available for computing rock permeability, such as arithmetic and harmonic averaging. The choice of method is case dependent, as each performs differently under varying conditions; see [7] for a detailed comparison. In *pycopm*, the default approach is to compute permeability in the $x$ and $y$ directions using the arithmetic average, while the $z$-direction permeability is calculated using the harmonic average. These defaults are particularly suitable for coarsening applied only in the vertical direction. As additional options, the permeability in coarsened cells $\mathbb{K}^*_{i^*,j^*,k^*}$ can be assigned using other operators such as min, max, or pore-volume-weighted mean of the permeabilities within the corresponding fine-scale cluster $\mathbb{K}_{i,j,k}$. For example, selecting the maximum permeability may be beneficial in history matching studies where the primary parameters being adjusted are relative permeabilities and capillary pressure (see [14] for a history matching study of an oil-gas field using *pycopm*).

For grids containing a large number of non-neighboring connections (such as faults) and inactive cells, upscaling transmissibilities is often a more effective and accurate approach than solely upscaling permeabilities. One limitation of this method is that permeability values cannot be directly used in history matching. Instead, transmissibility multipliers must be introduced, which increases the number of parameters to be matched and may disrupt workflows that rely on generating permeability fields from spatial correlations.

Two approaches to upscale transmissibilities are implemented in *pycopm*. The former computes the coarsened transmissibility

$$\mathbb{T}^*_{i^*,j^*,k^*} = [T^*_{i^*,j^*,k^* \to i^*+1,j^*,k^*}, T^*_{i^*-1,j^*,k^* \to i^*,j^*,k^*}, T^*_{i^*,j^*,k^* \to i^*,j^*+1,k^*}, T^*_{i^*,j^*-1,k^* \to i^*,j^*,k^*}, T^*_{i^*,j^*,k^* \to i^*,j^*,k^*+1}, T^*_{i^*,j^*,k^*-1 \to i^*,j^*,k^*}]$$

using the harmonic averaging along the transsmissibility direction and summing these values over the cell coarsened face. For example, for the transmissibility in the $z^+$ direction:

$$T^*_{i^*,j^*,k^* \to i^*,j^*,k^*+1} = \sum_{(i,j) \in \mathfrak{C}_{i^*,j^*,k^*}} \left( \sum_{(i,j,k) \to (i,j,k+1) \in \mathfrak{C}_{i^*,j^*,k^*}} \frac{1}{T_{i,j,k \to i,j,k+1}} \right)^{-1}.$$

In cases where coarsening is applied along a single direction, such as the vertical $z$-direction, the second method assigns the transmissibility on cell faces in that direction directly from the overlapping cell face values within the corresponding cluster, rather than computing a harmonic average. In preliminary testing, for input models with a large number of inactive cells, this approach has shown improved agreement with the original model compared to the harmonic averaging method.

For both transmissibility upscaling approaches, the values are scaled by the ratio of the effective cell face area in the original model to the corresponding coarse cell face area. This scaling is also applied to NNCs. In the coarsened model, the transmissibility of each NNC is computed as the sum of the corresponding values in the original model. This treatment is particularly important for preserving pressure communication across open faults that connect different formations.

Fig. 2 shows a comparison for coarsening permeabilities and transmissibilities in the corner-point grid described in Fig. 1 after 100 days of water injection.

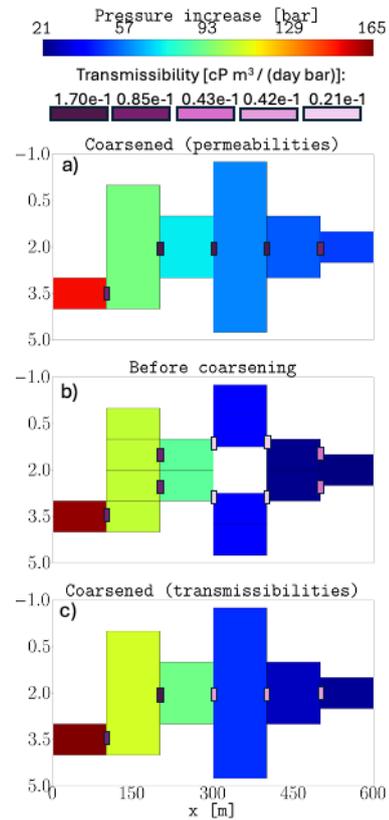

Figure 2: Pressure increase and transmissibility values for coarsening of only permeabilities and transmissibilities. The total pore volume in the three cases is 10,000 m³, and the pressures at the sensor (most right cell) are 47.3, 20.9, and 23.9 bar for a), b), and c) respectively.

Fig. 2a shows that coarsening based only on permeability values leads to higher transmissibility values in the coarsened model. This occurs because the overlapping cell face areas in the coarsened grid are larger than those in the original model. For example, in Fig. 2a, the first two transmissibility values (from left to right) match those of the original model in Fig. 2b. However, the remaining three values are overestimated, as the weighted-average permeability does not account for variations in the overlapping face areas. In contrast, when the model is coarsened using transmissibility values directly (Fig. 2c), the resulting transmissibilities match those of the original model.

## 2.2 Troll aquifer model

The Troll regional aquifer is located in the Horda Platform region on the Norwegian continental shelf, as illustrated in Fig. 3a. The model spans approximately 95 km in the west-east direction and 160 km in the south-north direction. Vertically, it extends from about 1 to 4 kms below sea level, as shown in Fig. 3b.

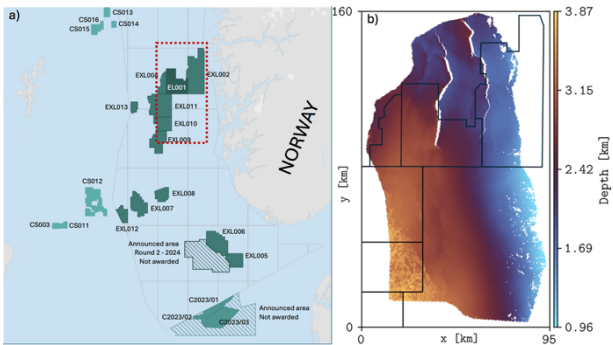

Figure 3: a) Map of the Norwegian continental shelf provided by the Norwegian Offshore Directorate, indicating the location of the Troll aquifer model. b) Spatial extent of the Troll aquifer model, showing its horizontal dimensions and vertical location.

The vertical thickness of the Troll aquifer model varies from 1 m to 852 m and comprises five distinct geological formations, as shown in Fig. 4.

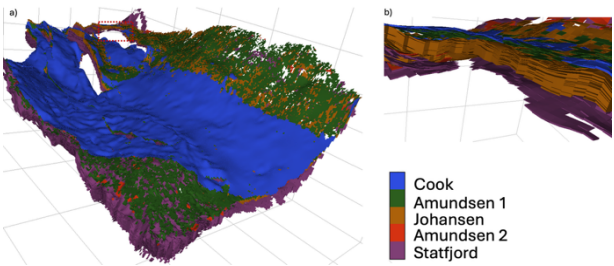

Figure 4: a) Troll aquifer model displaying the five modeled formations. b) Enlarged view of the northern red rectangular zone in a), highlighting the structural complexity of the grid.

The primary storage formations in the Troll aquifer model are Cook, Johansen, and Statfjord. In contrast, Amundsen 1 and 2 are partially eroded intervening shale layers, and subsequently contain a large number of disconnected and inactive cells. Fig. 5 shows the pore volumes associated with the main storage formations.

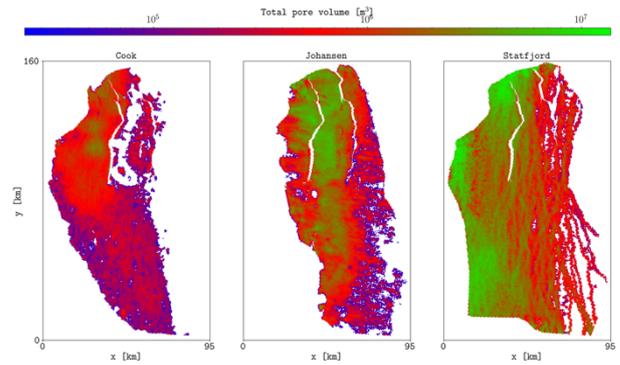

Figure 5: Total pore volume summed along the vertical direction for each of the main storage formations. Among these, Statfjord has the largest storage capacity, followed by Johansen and then Cook.

The Troll aquifer model contains numerous faults, as illustrated in Fig. 6a. In this study, the faults are considered to be open, allowing fluid communication between the geological formations. Fig. 6b highlights the faults that connect the three main storage formations, which are primarily located in the northwestern region of the model.

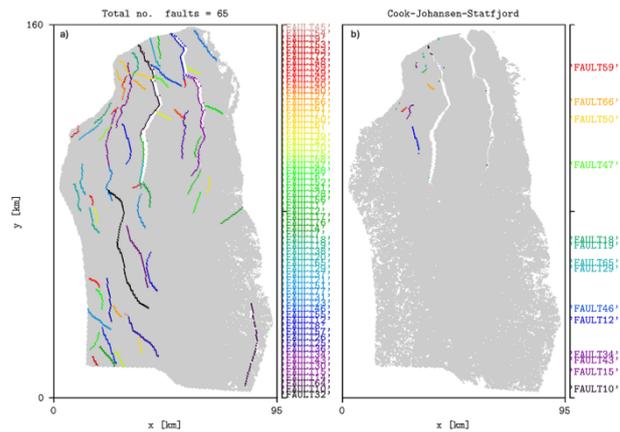

Figure 6: a) All faults in the Troll aquifer model. b) Faults establishing hydraulic connecting between the Cook, Johansen, and Statfjord formations.

Additional figures for the Troll regional model and further details on fluid and reservoir properties, including salinity, temperature, permeability, rock compressibility, and saturation functions, are provided in [15-17].

## 3. Results and Discussion

In this section, we apply the coarsening described in the previous section to the Troll aquifer model and use the coarsened model to conduct an optimization study. The aim of this study is to demonstrate the accuracy and efficiency of the coarsened model. Details on the commands and input decks required to reproduce both the preceding and subsequent results are available at https://github.com/cssr-tools/expreccs/tccs-13.

## 3.1 Troll aquifer coarsened model

The coarsening procedure is applied to each of the five geological formations, yielding a model with five cells in the vertical direction, as illustrated in Fig. 7.

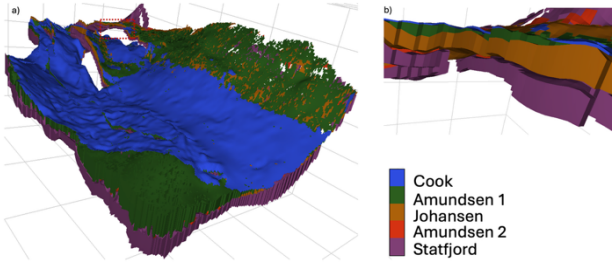

Figure 7: a) Coarsened Troll aquifer model displaying the five different formations. b) Enlarged view of the northern red rectangular zone in a).

Table 1 presents a comparison of model properties between the original and coarsened versions of the Troll aquifer model, including simulation times for a single run of the optimization study described in the next subsection.

Table 1: Summary of model properties.

|  | Troll model | Coarsened version |
|---|---|---|
| Number of cells | [181, 317, 217] | [181, 317, 5] |
| Active cells | 1,665,658 | 117,289 |
| Simulation time | 45 minutes | 1 minute |

Fig. 8 shows the pore-volume-weighted average pressure for both the original Troll model and its coarsened counterpart, including a 500-year post-injection monitoring period (for details on the set up for the simulations, see the next subsection). The results indicate that the coarsened model maintains strong agreement with the original model throughout the entire simulation period.

## 3.2 Optimization study

To demonstrate the value of the coarsened model, we consider a computationally intensive workflow that involves numerous model realizations, specifically an optimization study. It is worth noting that the coarsening approach is also applicable to other types of studies, such as sensitivity studies, history matching, and uncertainty quantification.

The objective of the optimization is to improve well placement by distributing pressure buildup more effectively throughout the system. The scenario involves injecting a total of 500 Mt over a 25-year period using 14 wells, each operating at a constant injection rate of 1.43 Mt per year. The wells are allowed to move freely in the horizontal plane but must be completed either they in the Cook-Johansen or Statfjord formations. Additionally, the maximum allowable pressure increase must not be exceeded (Fig. 9) [18]:

$$p_{lim} = \sigma z - p_0$$

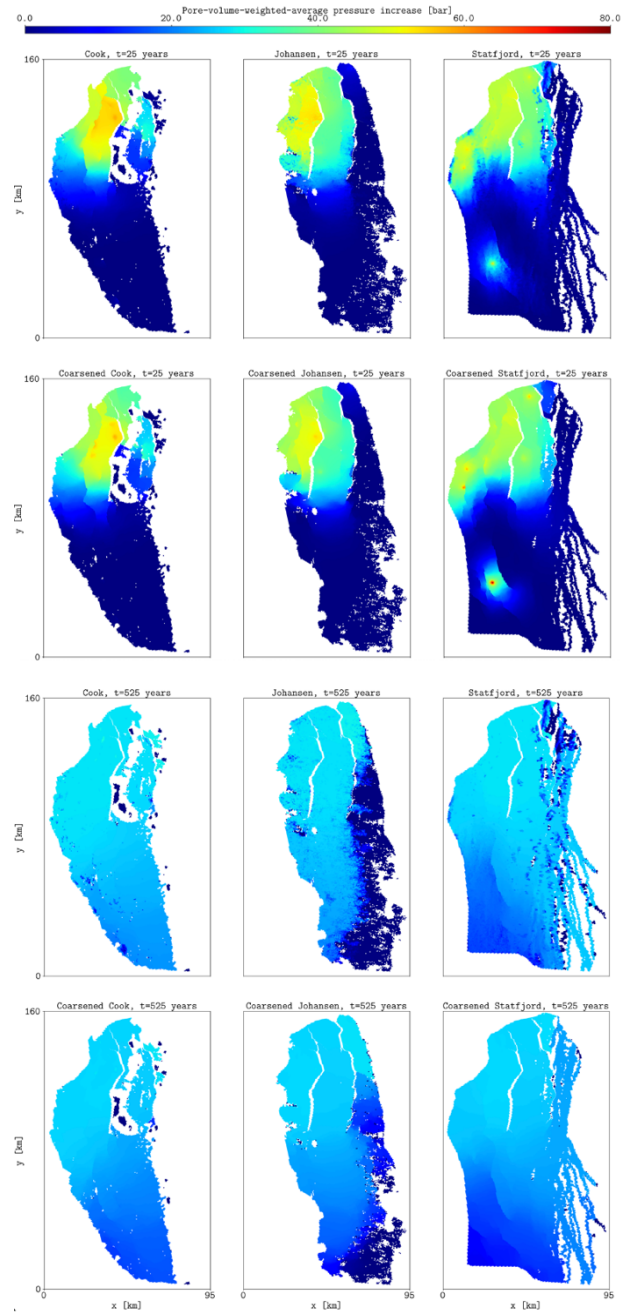

Fig. 8: Pore-volume-weighted averaged pressure at the end of the injection (25 years), and 500 years after the injection in the original Troll model and in the coarsened version.

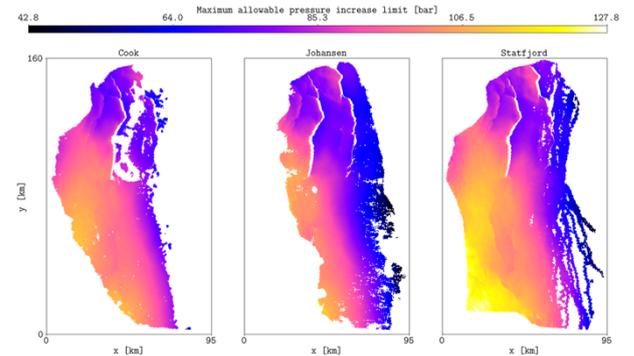

Figure 9: Maximum allowable pressure increase limit in each formation.

where $z$ is the depth of the most top cell face in the formation, $p_0$ the initial pressure, and $\sigma$ a model parameter equal to 0.134 [bar/m] [18].

We employ the differential evolution algorithm [19], as implemented in SciPy [20] and accessed through the Everest framework [21]. For further details on these widely used tools in optimization workflows, the reader is referred to their respective documentation. In practice, simulations are skipped if any proposed well is located entirely within inactive regions, i.e., outside the active model cells. To initialize the study, we use a feasible solution that satisfies all constraints; this initial configuration is shown in Fig. 10.

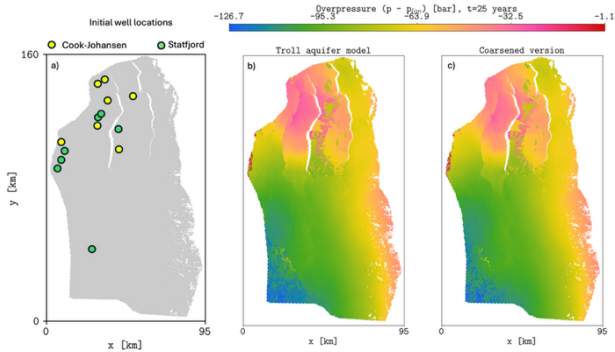

Figure 10: a) Initial well configuration, with seven wells completed in the Cook-Johansen formations and seven in the Statfjord formation. b) Overpressure in the Troll model and c) coarsened version (negative values indicate that the pressure remains below the specified maximum limit).

The objective function is:

$$\max_{wells_{x,y,l}} \left( min\left[1 - \frac{p}{p_{lim}}\right]\right)$$

where $l$ refers to the completion of the well, in this case Cook-Johansen or Statfjord formations. In words, this objective function maximizes the difference between the pressure limit and the pressure in the cells, by finding the cell(s) with the minimum difference.

Fig. 11a presents the evolution of the objective function across all 530 optimization steps, and Fig. 11b provides a breakdown of simulation outcomes, indicating the number of failed and successful runs. Failures occur when wells are placed in inactive regions, when the total stored mass is below 500 Mt, or when the pressure reached or exceeded the specified limit. The progression shown in Fig. 11 is notable, as the first significantly improved well configuration was identified around step 265, approximately halfway through the study. After this point, the number of executed simulations increased, and the study finished after 530 steps, with a total of 99,953 executed simulations.

Fig. 12 shows the improved well locations at the end of the study, along with the pressure distribution in both the original Troll model and the coarsened version. Compared to Fig. 10b and 10c, the high pressure observed in the easternmost cell is reduced by relocating the wells toward the north-central region.

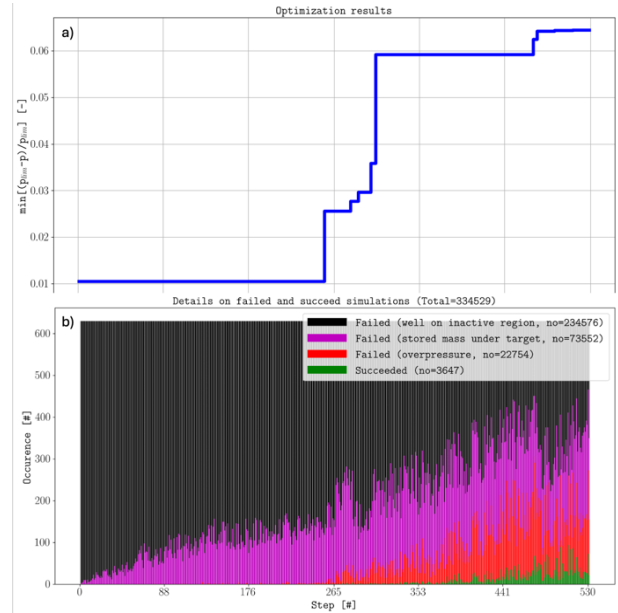

Figure 11: a) Objective function. b) Details on the simulations as a function of optimization steps.

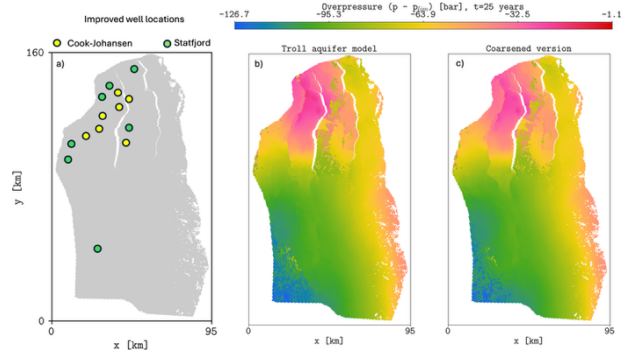

Fig. 12: a) Improved well locations. b) Overpressure in the Troll model and c) the coarsened version.

## 4. Conclusions

This paper presents the application of model reduction through grid and property coarsening in the Troll regional model. The coarsening method is based on transmissibility calculations, which account for both cell permeability and non-neighboring connections, such as offsets between cells used to represent faults. The accuracy and efficiency of the coarsened model is demonstrated through a well placement study, where the improved solution results in a reduced local pressure build-up in sensitive regions around faults.

The coarsened model enabled the execution of approximately 100,000 simulations within five days using a local server with 144 CPUs. In contrast, performing the same optimization study using the original Troll aquifer model would have required around seven months. While this approach is well suited for pressure-driven analyses, other quantities of interest, such as $CO_2$ migration, may require alternative model reduction techniques, including vertical equilibrium or sub-grid modeling.

The coarsening tool used in this study, *pycopm*, is openly available and can be applied to other models. In addition, all necessary details to reproduce the results presented in this paper are provided. This enables further exploration of alternative scenarios of interest, such as varying the number of injection wells, modifying the injection schedule, or changing the objective function.

## Acknowledgements

The authors are grateful for the financial support from the Research Council of Norway, Equinor ASA, A/S Norske Shell, Harbour Energy through project ExpReCCS (grant nr. 336294) and funding from the Centre for Sustainable Subsurface Resources (CSSR), grant nr. 331841, supported by the Research Council of Norway, research partners NORCE Research AS and the University of Bergen, and user partners Equinor ASA, Harbour Energy, Sumitomo Corporation, Earth Science Analytics, GCE Ocean Technology, and SLB Scandinavia. The authors also thank the Norwegian Offshore Directorate for providing access to the Troll Aquifer model.